\documentclass[preprints,review,accept]{mdpi}

\usepackage{revsymb}
\usepackage{amssymb}
\usepackage{siunitx}

\firstpage{1} 
\pubvolume{xx}
\issuenum{1}
\articlenumber{5}
\pubyear{2019}
\copyrightyear{2019}
\history{Received: date; Accepted: date; Published: date}

\Title{Probing vacuum polarization effects with high-intensity lasers}

\Author{Felix Karbstein $^{1,2,}$*}

\AuthorNames{Felix Karbstein}

\address{%
$^{1}$ \quad Helmholtz-Institut Jena, Fr\"obelstieg 3, 07743 Jena, Germany\\
$^{2}$ \quad Theoretisch-Physikalisches Institut, Abbe Center of Photonics, Friedrich-Schiller-Universit\"at Jena, Max-Wien-Platz 1, 07743 Jena, Germany
}

\corres{Correspondence: felix.karbstein@uni-jena.de}

\abstract{These notes provide a pedagogical introduction to the theoretical study of vacuum polarization effects in strong electromagnetic fields as provided by state-of-the-art high-intensity lasers.
Quantum vacuum fluctuations give rise to effective couplings between electromagnetic fields, thereby supplementing Maxwell's linear theory of classical electrodynamics with nonlinearities.
Resorting to a simplified laser pulse model allowing for explicit analytical insights, we demonstrate how to efficiently analyze all-optical signatures of these effective interactions in high-intensity laser experiments.
Moreover, we highlight several key features relevant for the accurate planning and quantitative theoretical analysis of quantum vacuum nonlinearities in the collision of high-intensity laser pulses.}

\keyword{External-field QED; Quantum Vacuum Nonlinearity; All-optical Signatures}

\begin{document}

\section{Introduction}

The quantum vacuum is not trivial and inert, but amounts to a complex state whose properties are fully determined by quantum fluctuations. 
These fluctuations comprise all existing particles, such that the quantum vacuum even constitutes a portal to new physics beyond the Standard Model of particle physics.
To allow for the principle measurement of these quantum vacuum fluctuations in experiment, the latter have to be excited by some external stimulus, such as strong macroscopic electromagnetic fields which couple directly to the charged particle sector.
Within the Standard Model, the leading effect arises from the effective coupling of the prescribed electromagnetic fields via a virtual electron-positron pair, and thus is governed by quantum electrodynamics (QED).
For these reasons, throughout these notes, we focus exclusively on the vacuum of quantum electrodynamics.

The present lecture notes, which can be considered as a natural continuation of the material presented in Ref.~\cite{Karbstein:2016hlj}, aim at providing the reader with a pedagogical introduction to the theoretical study of vacuum polarization effects in strong electromagnetic fields as provided by state-of-the-art high-intensity lasers.
In fact, they cover a broad range of aspects, from the theoretical foundations to the determination of the actual numbers of signal photons to be accessible in experiments using state-of-the-art technology.
The presented matters are organized as follows:

In Sec.~\ref{sec:HE} we briefly recall the physical processes described by the renowned Heisenberg-Euler effective Lagrangian, highlighting both manifestly nonperturbative and perturbative phenomena in Sec.~\ref{sec:nonpert} and Sec.~\ref{sec:pert}, respectively.
Here, we in particular argue that the leading effective interaction term arising from a perturbative weak-field expansion of the Heisenberg-Euler effective Lagrangian allows for the accurate theoretical description and study of all-optical signatures of QED vacuum nonlinearities accessible in the laboratory with present and near future technology on the one percent level.
All-optical signatures of quantum vacuum nonlinearity encompass the class of phenomena which are driven by electromagnetic fields of optical frequencies delivered by lasers, and result in purely photonic signals to be measured in experiment.

Subsequently, in Sec.~\ref{sec:j} we present a derivation of the differential number of signal photons induced in the effective vacuum-fluctuation-mediated nonlinear interaction between the driving electromagnetic fields.
The derivation presented here only requires basic knowledge of classical electrodynamics, and hence should be particularly suited for master and graduate students, as well as beginners in the field.

Section~\ref{sec:examples} details an explicit example of how all-optical signatures of quantum vacuum nonlinearity can be efficiently analyzed theoretically.
To this end, we limit ourselves to simple model field configurations, for which the necessary integrations can essentially be performed analytically.
After introducing a simplified laser pulse model in Sec.~\ref{sec:profiles}, we focus on the special scenario of two head-on colliding laser pulses in Sec.~\ref{sec:headon}.
In Sec.~\ref{sec:bothHIL} this scenario is then applied to the study of the prospective signal photon numbers attainable in the collision of two high-intensity laser pulses, and in Sec.~\ref{sec:biref} to the collision of a free-electron laser (FEL) and high-intensity laser pulse.

Finally, we end with conclusions and a very brief outlook in Sec.~\ref{sec:concls}.

\section{Heisenberg-Euler effective Lagrangian}\label{sec:HE}

The effective interactions of prescribed, macroscopic electromagnetic fields in the QED vacuum are governed by the renowned Heisenberg-Euler effective Lagrangian \cite{Heisenberg:1935qt}.
For slowly varying electromagnetic fields, i.e., fields which vary on spatial (temporal) scales much larger than the Compton wavelength (time) of the electron $\lambdabar_{\rm C}=\frac{\hbar}{m_ec}\simeq3.8\times10^{-13}\,{\rm m}$ ($\tau_{\rm C}=\frac{\lambdabar_{\rm C}}{c}\simeq1.3\times10^{-21}\,{\rm s}$), with electron mass $m_e$, the entire field dependence of the effective Lagrangian is encoded in the gauge and Lorentz invariant scalar quantities $\cal F$ and ${\cal G}^2$. The latter are defined as ${\cal F}=\frac{1}{4}F_{\mu\nu}F^{\mu\nu}=\frac{1}{2}(\vec{B}^2-\vec{E}^2)$ and ${\cal G}=\frac{1}{4}F_{\mu\nu}{}^\star F^{\mu\nu}=-\vec{B}\cdot\vec{E}$; ${}^\star F^{\mu\nu}$ denotes the dual field strength tensor and our metric convention is $g^{\mu\nu}={\rm diag}(-1,1,1,1)$. This dependence ensures the effective Lagrangian to be manifestly gauge and Lorentz invariant.
Correspondingly, in this limit we have

\begin{equation}
 {\cal L}_\text{eff}({\cal F},{\cal G}^2)={\cal L}_{\rm MW}+{\cal L}_\text{int}({\cal F},{\cal G}^2), \label{eq:Leff}
\end{equation}
where ${\cal L}_{\rm MW}=-{\cal F}$ denotes the classical Maxwell Lagrangian, and ${\cal L}_\text{int}({\cal F},{\cal G}^2)$ encodes nonlinear effective interactions of the electromagnetic field mediated by quantum vacuum fluctuations.
The effective Lagrangian~\eqref{eq:Leff} can be systematically derived from the microscopic theory of QED subjected to a prescribed macroscopic electromagnetic field by integrating out the quantum fields, i.e., the dynamical fermion and photon fields \cite{Dunne:2004nc,Gies:2016yaa}.
It can be organized in a perturbative loop expansion, with the leading effective interaction stemming from a one-loop diagram.
Higher-loop contributions are parametrically suppressed by powers of the fine-structure constant $\alpha=\frac{e^2}{4\pi\epsilon_0\hbar c}\simeq\frac{1}{137}$, where $e$ denotes the elementary charge.
In the remainder of these notes, we use the Heaviside-Lorentz System with $c=\hbar=1$. 
Upon restriction to the one-loop contribution, we obtain \cite{Heisenberg:1935qt,Schwinger:1951nm}

\begin{equation}
 {\cal L}_\text{int}({\cal F},{\cal G}^2)\simeq-\frac{1}{8\pi^2}\int_{0}^{\infty}\frac{{\rm d}T}{T^3}\,{\rm e}^{-m_e^2T}\biggl\{\frac{(eaT)(ebT)}{\tan(eaT)\tanh(ebT)}+\frac{1}{3}(eT)^2(a^2-b^2)-1\biggr\}\,, \label{eq:L1loop}
\end{equation}
where $a=\bigl(\sqrt{{\cal F}^2+{\cal G}^2}-{\cal F}\bigr)^{1/2}$, $b=\bigl(\sqrt{{\cal F}^2+{\cal G}^2}+{\cal F}\bigr)^{1/2}$, and the integration contour is implicitly assumed to lie slightly above the positive real $T$ axis, i.e., features a small positive imaginary part.
For a pedagogical introduction and derivation of the result~\eqref{eq:L1loop}, see Ref.~\cite{Karbstein:2016hlj}.
Equation~\eqref{eq:L1loop} forms the starting point of many theoretical studies and proposals aiming at the first experimental verification of QED vacuum nonlinearities in macroscopic electromagnetic fields in the laboratory; see the Reviews~\cite{Fradkin:1991zq,Dittrich:2000zu,Marklund:2006my,DiPiazza:2011tq,Dunne:2012vv,Battesti:2012hf,King:2015tba} and references therein.

\subsection{Manifestly non-perturbative physics}\label{sec:nonpert}

One of the most striking predictions of Eq.~\eqref{eq:L1loop} is the fact that it exhibits an imaginary part.
As the quantity ${\rm e}^{{\rm i}\int{\rm d}^4x\,{\cal L}_{\rm eff}(A)}=\langle0|0\rangle_A$ can be interpreted as vacuum persistence amplitude in the presence of the prescribed electromagnetic field (gauge potential $A\equiv A^\mu$) \cite{Schwinger:1951nm},

\begin{equation}
 P[A]=1-\bigl|\langle0|0\rangle_A\bigr|^2=1-{\rm e}^{-2\int{\rm d}^4x\,{\rm Im}\{{\cal L}_\text{int}(A)\}}
\end{equation}
amounts to the probability of the vacuum to decay, and $w=2{\rm Im}\{{\cal L}_\text{int}(A)\}$ can be interpreted as local decay rate.
Aiming at its explicit determination, we note that the imaginary part of ${\cal L}_\text{int}({\cal F},{\cal G}^2)$ can be expressed as

\begin{equation}
 {\rm Im}\bigl\{{\cal L}_\text{int}({\cal F},{\cal G}^2)\bigr\}=\frac{1}{2{\rm i}}\bigl[{\cal L}_\text{int}({\cal F},{\cal G}^2)-{\cal L}^*_\text{int}({\cal F},{\cal G}^2)\bigr]\,. \label{eq:ImL}
\end{equation}
As the integrand in Eq.~\eqref{eq:L1loop} is purely real-valued and the imaginary part stems only from the integration contour, the expression for ${\cal L}^*_\text{int}({\cal F},{\cal G}^2)$ amounts to the one for ${\cal L}_\text{int}({\cal F},{\cal G}^2)$ given in Eq.~\eqref{eq:L1loop}, with the integration contour now understood to lie slightly below the positive real $T$ axis.
By Cauchy's integral theorem, the difference in the square brackets on right side of Eq.~\eqref{eq:ImL} then corresponds to $-2\pi{\rm i}$ times the sum of the residues of the poles of the integrand lying on the positive real $T$ axis.
These poles arise from the inverse sine in the function $\frac{eaT}{\tan(eaT)}=\frac{1}{\sin(eaT)}eaT\cos(eaT)$, and are located at $eaT=n\pi$ with $n\in\mathbb{N}$. 
Taking into account that in the vicinity of the pole located at $eaT=n\pi$ we have $\frac{1}{\sin(eaT)}=\frac{(-1)^n}{ea}\frac{1}{T-\frac{n\pi}{ea}}+{\cal O}(T-\frac{n\pi}{ea})$, it is straightforwardly to infer that

\begin{equation}
 {\rm Res}\biggl\{\frac{eaT}{\tan(eaT)}f(T)\biggr\}\bigg|_{eaT=n\pi}=\frac{n\pi}{ea} f(\tfrac{n\pi}{ea})\,,
\end{equation}
where $f(T)$ is a function analytic at $eaT=n\pi$. 
With the help of this identity, it is easy to obtain the following result for the imaginary part of the effective Lagrangian \cite{Nikishov:1969tt},

\begin{equation}
 {\rm Im}\bigl\{{\cal L}_\text{int}({\cal F},{\cal G}^2)\bigr\}\simeq\frac{(ea)^2}{8\pi^3}\sum_{n=1}^\infty\,\frac{1}{n^2}\,{\rm e}^{-n\pi\frac{m_e^2}{ea}} \frac{\frac{eb}{ea}n\pi}{\tanh(\tfrac{eb}{ea}n\pi)}\,. \label{eq:ImL1loop}
\end{equation}
Obviously, the effective Lagrangian only develops an imaginary part for non-vanishing values of $a$.
Also note that this result is manifestly non-perturbative in the combined parameter $ea$, i.e., cannot be expanded in a power series around $ea=0$.
Making use of the identity $\frac{1}{\tanh(z)}=1+2\sum_{l=1}^\infty{\rm e}^{-2lz}$ for $z\in\mathbb{R}$, we arrive at an alternative representation of Eq.~\eqref{eq:ImL1loop},

\begin{equation}
 {\rm Im}\bigl\{{\cal L}_\text{int}({\cal F},{\cal G}^2)\bigr\}\simeq\frac{(ea)^2}{8\pi^3}\sum_{n=1}^\infty\,\frac{1}{n^2} \sum_{l=0}^\infty \frac{eb}{ea}n\pi\bigl(2-\delta_{0,l}\bigr)\,{\rm e}^{-n\pi\frac{m_e^2+2leb}{ea}}\,. \label{eq:ImL1loopA}
\end{equation}
Here, the summation index $l$ has an intuitive interpretation. It can be considered as labeling the $l$th Landau level in the magnetic-like field $b$; all levels apart from the zeroth one are doubly degenerate, and the effective mass squared of fermions in the $l$th Landau level is $m_e^2+2leb$.

For the special case of $b=0$, corresponding to a situation where ${\cal F}\leq0\,\leftrightarrow\,|\vec{B}|\leq|\vec{E}|$ and ${\cal G}=0\,\leftrightarrow\,\vec{B}\perp\vec{E}$, we have $a=\sqrt{2|{\cal F}|}=\sqrt{\vec{E}^2-\vec{B}^2}$ and

\begin{equation}
 {\rm Im}\bigl\{{\cal L}_\text{int}({\cal F},{\cal G}^2)\bigr\}\simeq\frac{(ea)^2}{8\pi^3}\sum_{n=1}^\infty\,\frac{1}{n^2}\,{\rm e}^{-n\pi\frac{m_e^2}{ea}}\,. \label{eq:ImL1loopE}
\end{equation}
In this case, one can always find a Lorentz frame in which $a$ amounts to a purely electric field, and Eq.~\eqref{eq:ImL1loopE} amounts to Schwinger's renowned formula describing pair production in a prescribed electric field \cite{Schwinger:1951nm}.

\subsection{Perturbative weak-field regime}\label{sec:pert}

On the other hand, resorting to the identity $\frac{z}{\tan(z)}=\sum_{n=0}^\infty(-1)^n\frac{{\cal B}_{2n}}{(2n)!}(2z)^{2n}$ for $|z|<\pi$ \cite{Gradshteyn}, noting that $\frac{z'}{\tanh(z')}=\frac{{\rm i}z'}{\tan({\rm i}z')}$, and making use of the Cauchy product, Eq.~\eqref{eq:L1loop} can be expanded as

\begin{align}
 {\cal L}_\text{int}({\cal F},{\cal G}^2)&\sim-\frac{1}{8\pi^2}\int_{0}^{\infty}\frac{{\rm d}T}{T^3}\,{\rm e}^{-m_e^2T}\biggl\{\sum_{n=2}^\infty\sum_{k=0}^n(-1)^k\frac{{\cal B}_{2k}}{(2k)!}\frac{{\cal B}_{2n-2k}}{(2n-2k)!}(2eaT)^{2k}(2ebT)^{2(n-k)}\biggr\} \nonumber\\
 &\sim-\frac{m_e^4}{8\pi^2}\sum_{n=2}^\infty\sum_{k=0}^n(-1)^k(2n-3)! \frac{{\cal B}_{2k}}{(2k)!}\frac{{\cal B}_{2n-2k}}{(2n-2k)!}\Bigl(\frac{2ea}{m_e^2}\Bigr)^{2k}\Bigl(\frac{2eb}{m_e^2}\Bigr)^{2(n-k)} \label{eq:L1loopass}
\end{align}
where the integration over the propertime variable $T$ could be performed trivially as $\int_0^\infty\frac{{\rm d}T}{T^3}\,{\rm e}^{-m_e^2T}T^{2n}=(-\partial_{m_e^2})^{2n-3}\int_0^\infty{\rm d}T\,{\rm e}^{-m_e^2T}=(-\partial_{m_e^2})^{2n-3}\frac{1}{m_e^2}=m_e^4\frac{(2n-3)!}{(m_e^2)^{2n}}$.
Note, that the Bernoulli numbers appearing in this expansion can be expressed as ${\cal B}_{2n}=\frac{(-1)^{n+1}2(2n)!}{(2\pi)^{2n}}\zeta(2n)$ \cite{Arfken} in terms of the Riemann zeta function, $\zeta(n)=\sum_{k=1}^\infty\frac{1}{k^n}$ for $n\in\mathbb{N}$ and $\zeta(0)=-\frac{1}{2}$. Clearly, $\lim_{n\to\infty}\zeta(n)=1$.

From Eq.~\eqref{eq:L1loopass} it is obvious that ${\cal L}_\text{int}({\cal F},{\cal G}^2)$ accounts for all orders in a perturbative expansion in the strength of the applied electromagnetic field.
In accordance with {\it Furry's theorem} \cite{Furry} (charge conjugation invariance of QED), Eq.~\eqref{eq:L1loopass} is even in the elementary charge $e$. 
As to be expected, this perturbative expansion is completely insensitive to the manifestly nonperturbative imaginary part isolated in Sec.~\ref{sec:nonpert} above.
For fields fulfilling $\{\frac{ea}{m_e^2},\frac{eb}{m_e^2}\}\ll1$, already a truncation of Eq.~\eqref{eq:L1loopass} to the leading term should allow for reliable insights; the imaginary part of ${\cal L}_\text{int}({\cal F},{\cal G}^2)$ determined above is exponentially suppressed in this limit.
The latter criterion is equivalent to $\{\frac{E}{E_{\rm cr}},\frac{B}{B_{\rm cr}}\}\ll1$, with the so-called critical electric and magnetic fields defined as $E_\text{cr}=\frac{m_e^2c^3}{e\hbar}\simeq 1.3 \times 10^{18}\,\frac{\rm V}{\rm m}$ and $B_\text{cr}=\frac{E_\text{cr}}{c} \simeq 4 \times 10^{9}\,{\rm T}$, respectively. All macroscopic electromagnetic fields presently available in the laboratory, particularly also those generated by the most advanced high-intensity laser facilities, such as CILEX \cite{CILEX}, CoReLS \cite{CoReLS}, ELI \cite{ELI} and SG-II \cite{SG-II}, perfectly fulfill this criterion.
From a more conceptual point of view, it is however important to note that Eq.~\eqref{eq:L1loopass} constitutes an asymptotic expansion: the Bernoulli numbers ${\cal B}_{2n}$ alternate in sign and diverge factorially fast in magnitude.

\begin{figure}
\centering
 \includegraphics[width=0.25\linewidth]{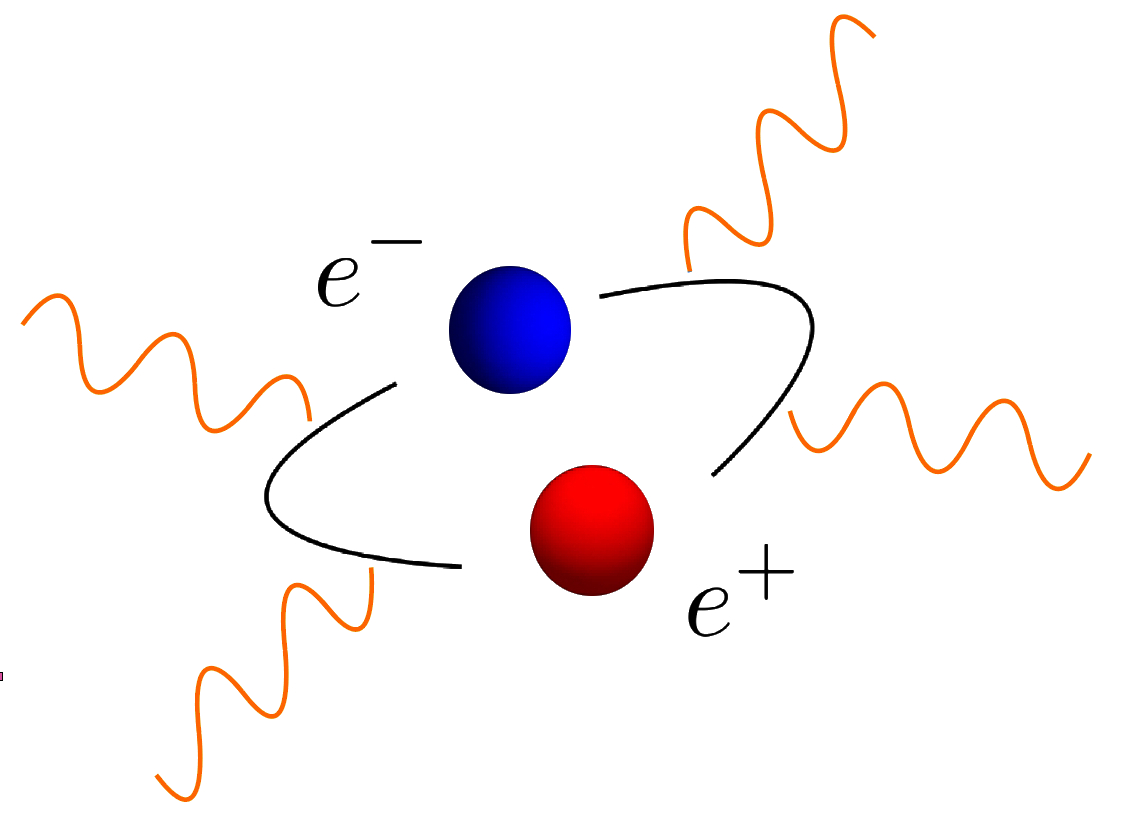}
 \caption{Illustration of the effective interaction between four electromagnetic fields (wiggly orange lines) mediated by a virtual electron-positron fluctuation in the QED vacuum.}
 \label{fig:4gamma}
\end{figure}

Formally counting $\frac{ea}{m_e^2}\sim \frac{eb}{m_e^2}\sim\epsilon$, where $\epsilon$ denotes a dimensionless expansion parameter, the leading term amounts to the contribution of ${\cal O}(\epsilon^4)$.
It is given by

\begin{equation}
 {\cal L}_\text{int}({\cal F},{\cal G}^2)\simeq\frac{m_e^4}{360\pi^2}\Bigl(\frac{e}{m_e^2}\Bigr)^4\bigl(a^4+5a^2b^2+b^4\bigr) \label{eq:L1loopleading}
 =\frac{m_e^4}{360\pi^2}\Bigl(\frac{e}{m_e^2}\Bigr)^4\bigl(4{\cal F}^2+7{\cal G}^2\bigr)\,.
\end{equation}
See Fig.~\ref{fig:4gamma} for an illustration of the effective interaction in Eq.~\eqref{eq:L1loopleading}.
Higher order corrections to Eq.~\eqref{eq:L1loopleading} are parametrically suppressed by powers of the dimensionless parameters $\frac{E}{E_{\rm cr}}$ and $\frac{B}{B_{\rm cr}}$, measuring the strengths of the prescribed electric and magnetic fields in units of the respective critical field, as well as the fine-structure constant $\alpha$.

Another important point to clarify is the applicability of the Heisenberg-Euler effective Lagrangian~\eqref{eq:Leff}, which has formally been derived for the case of constant electromagnetic fields, to manifestly inhomogeneous fields.
As already noted above, in QED the important reference scales the spatial and temporal variations of the electromagnetic field are to be compared with are the Compton wavelength and time of the electron, respectively.
This comes about as follows \cite{Karbstein:2015cpa}: in constant electromagnetic fields the requirement of gauge and Lorentz invariance render the effective Lagrangian a function of the scalar quantities $\mathcal F$ and ${\mathcal G}^2$ only.
For inhomogeneous background fields, however, additional gauge and Lorentz invariant building blocks involving derivatives of the field strength tensor become available.
The deviations from the constant field limit can be accounted for with derivative terms $\sim\partial_\alpha F^{\mu\nu}$.
If the typical spatial and temporal scales of variation of the inhomogeneous field are $L$ and $T=L/c$, respectively, derivatives effectively translate into multiplications with $\frac{1}{L}$ to be rendered dimensionless by the electron mass, constituting the only dimensionful parameter of low-energy QED.
In turn, the above results formally derived for constant fields are also applicable for slowly varying inhomogeneous fields fulfilling $\frac{1}{m_eL}\ll1$, or equivalently $L\gg\lambdabar_{\rm C}$ and $T\gg\tau_{\rm C}$.
The effective Lagrangian is a scalar quantity, and scalar quantities made up of combinations of $F^{\mu\nu}$, ${}^\star F^{\mu\nu}$ and derivatives thereof involve an even number of derivatives.
This, in particular implies that the above constant-field result allows for the reliable description of inhomogeneous electromagnetic fields up to corrections of ${\cal O}\bigl((\frac{\lambdabar_{\rm C}}{L})^2\bigr)$.
In the literature, neglecting these corrections is often referred to as resorting to a locally constant field approximation (LCFA).

Finally, we emphasize that Eq.~\eqref{eq:L1loopleading} serves as an ideal starting point for the discussion of all-optical signatures of QED vacuum nonlinearity in strong electromagnetic fields accessible with state-of-the-art and near-future experimental technology, detailed below.
To this end, first recall that $\alpha\simeq\frac{1}{137}\ll1$ and the peak field strengths reached by state-of-the-art high-intensity lasers, $E={\cal O}(10^{14})\frac{\rm V}{\rm m}$ and $B={\cal O}(10^{6}){\rm T}$, clearly fulfill $\{\frac{E}{E_{\rm cr}},\frac{B}{B_{\rm cr}}\}\ll1$. Second, note that the slowly-varying field criterion $(\frac{\lambdabar_{\rm C}}{L})^2\ll1$ is safely met for laser photon energies $\omega$ up to those delivered by latest generation FELs, reaching $\omega\simeq25\,{\rm keV}$ \cite{LCLS-II,XFEL}.
For these reasons, Eq.~\eqref{eq:L1loopleading} should even allow for the accurate theoretical analysis of all-optical probes of QED vacuum nonlinearities on the one percent level.

\section{Classical derivation of the differential signal photon number}\label{sec:j}

In this section, we aim at obtaining the differential number of signal photons arising from the vacuum-fluctuation-mediated nonlinear effective interaction of the driving pump fields \cite{Galtsov:1971xm,Karbstein:2014fva}.
This will serve as the central input for studies of all-optical signatures of QED vacuum nonlinearity in strong electromagnetic fields, as provided by high-intensity laser systems.
See Sec.~\ref{sec:examples} for the discussion of exemplary applications.
The derivation presented here only requires basic knowledge of classical electrodynamics, and hence should be easily accessible for readers without any background in quantum field theory, and specifically QED. 

To this end, we decompose $F^{\mu\nu}\to F^{\mu\nu}+f^{\mu\nu}$ into an applied electromagnetic field $F^{\mu\nu}$ and a signal photon field $f^{\mu\nu}$, with the former typically being much stronger than the latter. Accounting for terms up to linear order in the signal photon field, we obtain

\begin{equation}
 {\cal L}_{\rm int}(F+f)={\cal L}_{\rm int}(F)+\frac{\partial{\cal L}_{\rm int}(F)}{\partial F^{\mu\nu}}f^{\mu\nu}+{\cal O}(f^2)\,. \label{eq:Lf}
\end{equation}
The neglected higher-order contributions encode processes giving rise to more than one outgoing signal photon, which are typically suppressed relatively to single photon emission processes.
A prominent nonlinear QED signature characterized by two signal photons in the final state is photon splitting \cite{BialynickaBirula:1970vy,Adler:1970gg,Adler:1971wn,Lee:1998hu,Papanyan:1971cv,Stoneham:1979,Baier:1986cv,Adler:1996cja,DiPiazza:2007yx}.
On the other hand, prospective optical signatures such as vacuum birefringence \cite{Toll:1952,Baier,BialynickaBirula:1970vy,Adler:1971wn,Aleksandrov:1985,Kotkin:1996nf,Heinzl:2006xc,DiPiazza:2006pr,Dinu:2013gaa,Nakamiya:2015pde,Karbstein:2015xra,Ilderton:2016khs,King:2016jnl,Schlenvoigt:2016,Karbstein:2016lby,Bragin:2017yau}, quantum reflection \cite{Gies:2013yxa,Gies:2014wsa}, photon merging \cite{Yakovlev:1966,DiPiazza:2007cu,Gies:2014jia,Gies:2016czm}, generic photon scattering phenomena \cite{Rozanov:1993,Mckenna:1963,Varfolomeev:1966,Moulin:2002ya,Lundstrom:2005za,Tommasini:2010fb,Hatsagortsyan:2011,King:2012aw,King:2013am,Gies:2017ygp,Gies:2017ezf,King:2018wtn,Karbstein:2019dxo}, and higher-harmonic generation \cite{BialynickaBirula:1981,Kaplan:2000aa,Valluri:2003sp,Fedotov:2006ii,Marklund:2006my,DiPiazza:2005jc,King:2014vha,Bohl:2015uba,Kadlecova:2018vty} can be recast as single signal photon emission processes.
Upon explicit insertion of the definition of the field strength tensor of the signal photon field $f^{\mu\nu}=\partial^\mu a^\nu-\partial^\nu a^\mu$, Eq.~\eqref{eq:Lf} can be rewritten as

\begin{align}
 {\cal L}_{\rm int}(F+f)&={\cal L}_{\rm int}(F)+2\frac{\partial{\cal L}_{\rm int}(F)}{\partial F^{\mu\nu}}\partial^\mu a^\nu+{\cal O}(a^2) \nonumber\\
 &={\cal L}_{\rm int}(F)-\underbrace{2\partial^\mu\Bigl(\frac{\partial{\cal L}_{\rm int}(F)}{\partial F^{\mu\nu}}\Bigr)}_{=:j_\nu(F)} a^\nu+{\cal O}(a^2)\,, \label{eq:jnu}
\end{align}
where we made use of partial integration to shift away the derivative $\partial^\mu$ from the signal photon field $a^\nu$ in the second step. In typically considered scenarios, the driving classical electromagnetic fields are sufficiently strong only in a limited space-time volume, such that ${\cal L}_{\rm int}(F)$ can safely be assumed to vanish identically far outside the interaction region.
Moreover, we introduced the signal photon current $j_\nu(F)$ sourcing the signal photon field $a^\nu$.

In turn, the effective theory governing the dynamics of the single signal emission process is

\begin{equation}
 {\cal L}_\gamma(a|F)=-\frac{1}{4}f_{\mu\nu}f^{\mu\nu}-j_\nu\,a^\nu
\end{equation}
with $j_\nu\equiv j_\nu(F)$.
The associated equations of motion follow from the Euler-Lagrange equations, $\frac{\partial{\cal L}_\gamma}{\partial a_\nu}-2\partial_\mu\frac{\partial{\cal L}_\gamma}{\partial f_{\mu\nu}}=0$, and are given by

\begin{equation}
 \partial_\mu f^{\mu\nu}=j^\nu(x)\,. \label{eq:EoM}
\end{equation}
In Lorenz gauge, $\partial_\mu a^\mu=0$, which we adopt subsequently, Eq.~\eqref{eq:EoM} becomes

\begin{equation}
\Box\,a^\nu = j^\nu(x)\,,
\end{equation}
with $\Box=\partial_\mu\partial^\mu$ denoting the d'Alembert operator.
The solution of this equation, describing outgoing signal photons sourced by the current $j^\nu(F)$ is 

\begin{equation}
 a^\nu(x)=\int{\rm d}^4x'\,G_{\rm R}(x,x')\,j^\nu(x')\,, \label{eq:a}
\end{equation}
where $G_{\rm R}$ is the retarded Green's function of the d'Alembert operator.
A useful integral representation of this Green's function is (cf., e.g., Ref.~\cite{Gies:2014wsa})
\begin{equation}
 G_{\rm R}(x,x')=\Theta(t-t')\,\int\frac{{\rm d}^3k}{(2\pi)^3}\frac{1}{2k^0}\Bigl(\frac{1}{\rm i}\,{\rm e}^{{\rm i}k(x-x')}+\text{c.c.}\Bigr)\bigg|_{k^0=|\vec{k}|}\,,
\end{equation}
with Heaviside function $\Theta(.)$. The shorthand c.c. stands for complex conjugate.

In our calculation we will assume the detection of the signal photons to take place at asymptotic times $t$ well after the interaction, such that $t-t'>0$ is always fulfilled.
As the signal photon current receives substantial contributions only within the interaction region and decays rapidly to zero outside (cf. above), we can nevertheless formally extend the integration over $t'$ to $+\infty$.
Correspondingly, Eq.~\eqref{eq:a} results in

\begin{equation}
 a^\nu(x)=\int\frac{{\rm d}^3k}{(2\pi)^3}\frac{1}{2k^0}\Bigl(\frac{{\rm e}^{{\rm i}kx}}{\rm i}\,\underbrace{\int{\rm d}^4x'{\rm e}^{-{\rm i}kx'}j^\nu(x')}_{=j^\nu(k)}+\text{c.c.}\Bigr)\bigg|_{k^0=|\vec{k}|}\,,
\end{equation}
which for manifestly real-valued $j^\nu(x)$ can be compactly written as

\begin{equation}
 a^\nu(x)={\rm Re}\int\frac{{\rm d}^3k}{(2\pi)^3}\frac{1}{k^0}\frac{{\rm e}^{{\rm i}kx}}{\rm i}\,j^\nu(k)\bigg|_{k^0=|\vec{k}|}\,.
\end{equation}
The associated electric $\vec{e}(x)$ and magnetic $\vec{b}(x)$ fields are

\begin{align}
 \vec{e}(x)&=-\vec{\nabla}a^0(x)-\partial_t\vec{a}(x)={\rm Re}\int\frac{{\rm d}^3k}{(2\pi)^3}\,{\rm e}^{{\rm i}kx}\underbrace{\Bigl(\vec{j}(k)-\hat{\vec{k}}\,j^0(k)\Bigr)}_{=:\vec{e}(k)}\bigg|_{k^0=|\vec{k}|}\,, \nonumber\\
 \vec{b}(x)&=\vec{\nabla}\times\vec{a}(x)={\rm Re}\int\frac{{\rm d}^3k}{(2\pi)^3}\,{\rm e}^{{\rm i}kx}\underbrace{\Bigl(\hat{\vec{k}}\times \vec{j}(k)\Bigr)}_{=:\vec{b}(k)}\bigg|_{k^0=|\vec{k}|}\,. \label{eq:eb}
\end{align}
Equation~\eqref{eq:eb} directly implies that $\vec{k}\cdot\vec{e}(k)\big|_{k^0=|\vec{k}|}=\bigl(\vec{k}\cdot\vec{j}(k)-|\vec{k}|\,j^0(k)\bigr)\big|_{k^0=|\vec{k}|}=k_\mu j^\mu(k)\big|_{k^0=|\vec{k}|}=0$, where we made use of the Ward identity for the photon current $k_\mu j^\mu(k)=0$ in the last step. This Ward identity follows straightforwardly from the definition of the current in Eq.~\eqref{eq:jnu}.
Moreover, $\hat{\vec{k}}\times\vec{e}(k)=\hat{\vec{k}}\times\vec{j}(k)=\vec{b}(k)$, such that both the electric and magnetic fields are transverse, and $|\vec{e}(k)|=|\vec{b}(k)|=|\vec{j}(k)|$.

The energy put into the signal photon field is

\begin{equation}
 W=\frac{1}{2}\int{\rm d}^3x\,\bigl(|\vec{e}(x)|^2+|\vec{b}(x)|^2\bigr)\,. \label{eq:W}
\end{equation}
With the help of Eq.~\eqref{eq:eb} we can write the first term in Eq.~\eqref{eq:W} as

\begin{align}
 \int{\rm d}^3x\,|\vec{e}(x)|^2&=\frac{1}{4}\int{\rm d}^3x\int\frac{{\rm d}^3k}{(2\pi)^3}\int\frac{{\rm d}^3k'}{(2\pi)^3}
 \Bigl({\rm e}^{{\rm i}kx}\,\vec{e}(k)+{\rm e}^{-{\rm i}kx}\,\vec{e}^{\,*}(k)\Bigr) \nonumber\\ 
 &\hspace*{4.14cm}\times\Bigl({\rm e}^{{\rm i}k'x}\,\vec{e}(k')+{\rm e}^{-{\rm i}k'x}\,\vec{e}^{\,*}(k')\Bigr)\bigg|_{k^0=|\vec{k}|,k'^0=|\vec{k}'|}\nonumber\\
 &=\frac{1}{4}\int\frac{{\rm d}^3k}{(2\pi)^3}\Bigl(2\,\vec{e}(k)\vec{e}^{\,*}(k)+{\rm e}^{2{\rm i}k^0t}\,\vec{e}(k)\vec{e}(-k)+{\rm e}^{-2{\rm i}k^0t}\,\vec{e}^{\,*}(k)\vec{e}^{\,*}(-k)\Bigr)\bigg|_{k^0=|\vec{k}|}\,,
\end{align}
where ${}^*$ denotes complex conjugation.

Doing the same manipulations for the second term in Eq.~\eqref{eq:W} and using the identities $\vec{b}(k)\cdot\vec{b}(-k)=\bigl(\hat{\vec{k}}\times\vec{e}(k)\bigr)\bigl(-\hat{\vec{k}}\times\vec{e}(-k)\bigr)=-\vec{e}(k)\cdot\vec{e}(-k)$ as well as $\vec{b}^{\,*}(k)\cdot\vec{b}(-k)=-\vec{e}^{\,*}(k)\cdot\vec{e}(-k)$, the signal photon energy can be expressed as

\begin{equation}
 W=\frac{1}{4}\int\frac{{\rm d}^3k}{(2\pi)^3}\bigl(|\vec{e}(k)|^2+|\vec{b}(k)|^2\bigr)\bigg|_{k^0=|\vec{k}|}
 =\frac{1}{2}\int\frac{{\rm d}^3k}{(2\pi)^3} |\vec{e}(k)|^2\bigg|_{k^0=|\vec{k}|}\,. \label{eq:Wk}
\end{equation}
Making use of the fact that the electric field can be spanned by two orthogonal polarization vectors $\vec{\epsilon}_p(k)$ transverse to $\vec{k}$, with $p\in\{1,2\}$, we obtain

\begin{equation}
  |\vec{e}(k)|^2 = \sum_{p=\{1,2\}}|\vec{\epsilon}_p(k)\cdot\vec{e}(k)|^2 = \sum_{p=\{1,2\}}|\vec{\epsilon}_p(k)\cdot\vec{j}(k)|^2 \,,
\end{equation}
and thus

\begin{equation}
 W=\sum_{p=\{1,2\}}\int\frac{{\rm d}^3k}{(2\pi)^3}\,\frac{1}{2}|\vec{\epsilon}_p(k)\cdot\vec{j}(k)|^2 \bigg|_{k^0=|\vec{k}|}\,.
\end{equation}
From this expression, we infer that the differential field energy put into mode $p$ is

\begin{equation}
 {\rm d}^3W_p=\frac{{\rm d}^3k}{(2\pi)^3}\,\frac{1}{2}|\vec{\epsilon}_p(k)\cdot\vec{j}(k)|^2 \bigg|_{k^0=|\vec{k}|}\,. \label{eq:dWp}
\end{equation}
The differential number of signal photons of energy $k^0=|\vec{k}|$ emitted into the solid angle interval ${\rm d}^2\Omega$, with ${\rm d}^3k=\vec{k}^2{\rm d}k\,{\rm d}^2\Omega$, follows from Eq.~\eqref{eq:dWp} upon division by the photon energy, yielding

\begin{equation}
 {\rm d}^3N_p=\frac{{\rm d}^3k}{(2\pi)^3}\,\frac{1}{2k^0}|\vec{\epsilon}_p(k)\cdot\vec{j}(k)|^2 \bigg|_{k^0=|\vec{k}|}\,. \label{eq:dNp}
\end{equation}
For definiteness, throughout these notes we use conventions where $\hat{\vec{k}}\times\vec{\epsilon}_1(k)=\vec{\epsilon}_2(k)$, and thus $\hat{\vec{k}}\times\vec{\epsilon}_p(k)=\vec{\epsilon}_{p+1}(k)$, with $\vec{\epsilon}_3(k)\equiv-\vec{\epsilon}_1(k)$.

Recall, that in our case the signal photon current in momentum space follows upon Fourier transform of the expression given in Eq.~\eqref{eq:jnu}. In turn, we have

\begin{equation}
 j_\nu(k)=\int{\rm d}^4x\,{\rm e}^{-{\rm i}kx}\,2\partial^\mu\Bigl(\frac{\partial{\cal L}_{\rm int}(F)}{\partial F^{\mu\nu}}\Bigr)
 =2{\rm i}k^\mu\int{\rm d}^4x\,{\rm e}^{-{\rm i}kx}\,\frac{\partial{\cal L}_{\rm int}(F)}{\partial F^{\mu\nu}}\,,
\end{equation}
which for ${\cal L}_{\rm int}(F)={\cal L}_{\rm int}({\cal F},{\cal G}^2)$ as is assumed here can be expressed as 

\begin{equation}
 j_\nu(k)={\rm i}k^\mu\int{\rm d}^4x\,{\rm e}^{-{\rm i}kx} \Bigl(F_{\mu\nu}\frac{\partial{\cal L}_{\rm int}}{\partial{\cal F}}+{}^\star F_{\mu\nu}\frac{\partial{\cal L}_{\rm int}}{\partial{\cal G}}\Bigr)\,. \label{eq:jk}
\end{equation}
Taking into account the following identities,

\begin{align}
 k^\mu F_{\mu\nu}\big|_{k^0=|\vec{k}|}&=k^0\bigl(\hat{\vec{k}}\cdot\vec{E},-\hat{\vec{k}}\times\vec{B}-\vec{E}\bigr)\,, \nonumber\\
 k^\mu{}^\star F_{\mu\nu}\big|_{k^0=|\vec{k}|}&=k^0\bigl(\hat{\vec{k}}\cdot\vec{B},\hat{\vec{k}}\times\vec{E}-\vec{B}\bigr)\,,
\end{align}
the amplitude entering Eq.~\eqref{eq:dNp} can finally be written as,

\begin{equation}
 \vec{\epsilon}_p(k)\cdot\vec{j}(k)\big|_{k^0=|\vec{k}|}={\rm i}k^0\int{\rm d}^4x\,{\rm e}^{-{\rm i}kx} \bigl[\vec{\epsilon}_p(k)\cdot\vec{P}-\vec{\epsilon}_{p+1}(k)\cdot\vec{M}\bigr]\Big|_{k^0=|\vec{k}|}\,, \label{eq:epj}
\end{equation}
with

\begin{align}
 \vec{P}&:=\frac{\partial{\cal L}_{\rm int}}{\partial\vec{E}}=-\vec{E}\frac{\partial{\cal L}_{\rm int}}{\partial{\cal F}}-\vec{B}\frac{\partial{\cal L}_{\rm int}}{\partial{\cal G}}\,, \nonumber\\
 \vec{M}&:=-\frac{\partial{\cal L}_{\rm int}}{\partial\vec{B}}=-\vec{B}\frac{\partial{\cal L}_{\rm int}}{\partial{\cal F}}+\vec{E}\frac{\partial{\cal L}_{\rm int}}{\partial{\cal G}}\,, \label{eq:PM}
\end{align}
denoting the polarization $\vec{P}$ and magnetization $\vec{M}$ vectors of the quantum vacuum in the presence of the prescribed electromagnetic field.
Taking into account only the leading QED vacuum nonlinearity~\eqref{eq:L1loopleading} in perturbatively weak fields, Eqs.~\eqref{eq:PM} can be compactly written as

\begin{align}
 \vec{P}&\simeq\sqrt{\frac{\alpha}{\pi}}\frac{m_e^2}{90\pi}\Bigl(\frac{e}{m_e^2}\Bigr)^3\bigl[-2\vec{E}(\vec{B}^2-\vec{E}^2)+7(\vec{B}\cdot\vec{E})\vec{B}\,\bigr]\,, \nonumber\\
 \vec{M}&\simeq\sqrt{\frac{\alpha}{\pi}}\frac{m_e^2}{90\pi}\Bigl(\frac{e}{m_e^2}\Bigr)^3\bigl[-2\vec{B}(\vec{B}^2-\vec{E}^2)-7(\vec{B}\cdot\vec{E})\vec{E}\,\bigr]\,, \label{eq:PMleading}
\end{align}
where we expressed the quantities $\cal F$ and $\cal G$ in terms of the electric and magnetic fields; see the definitions above Eq.~\eqref{eq:Leff}.
Hence, the determination of the differential signal photon number~\eqref{eq:dNp} for experimentally viable field configurations in the laboratory boils down to performing four-dimensional Fourier transforms of the polarization and magnetization vectors~\eqref{eq:PMleading} of the quantum vacuum in position space.
While Eq.~\eqref{eq:dNp} can be readily evaluated for generic, experimentally realistic field configurations fulfilling Maxwell's equations in vacuo exactly \cite{Blinne:2018nbd}, in the subsequent considerations we will limit ourselves to simple model field configurations, for which the necessary integrations can essentially be performed analytically.

For an alternative derivation of the same result making use of a quantum Fock space formulation and asymptotic states, see Refs.~\cite{Karbstein:2014fva,Gies:2017ygp}.
Apart from a normalization factor, in this context Eq.~\eqref{eq:epj} amounts to the zero-to-single signal photon emission amplitude in the quantum vacuum subjected to the driving electromagnetic fields.
In particular using the Fock space formulation, a generalization of the vacuum emission approach to states featuring more than one signal photon in the final state is straightforward.

Also note that the derivation of the differential signal photon number does not rely on the LCFA. Modifying the derivation of the signal photon current accordingly \cite{Gies:2017ygp}, it can be readily employed for field configurations not amenable to an LCFA. See Ref.~\cite{Aleksandrov:2019irn} for an explicit example of a calculation using the vacuum emission picture beyond the LCFA.

\section{All-optical signatures of quantum vacuum nonlinearity}\label{sec:examples}

In the following discussion, we assume the electromagnetic fields driving the signal photon emission process to be provided by state-of-the-art high-intensity laser systems. The latter are characterized by an oscillation wavelength $\lambda_l$ (frequency $\omega_l=\frac{2\pi}{\lambda_l}$), and deliver pulses of energy $W_l$ and duration $\tau_l$ at a certain repetition rate; the index labels the parameters of laser system $l$.
For simplicity and in analogy to plane waves, we assume the macroscopic electric $\vec{E}_l$ and magnetic $\vec{B}_l$ fields constituting a given laser pulse to be  equal in magnitude and strictly transverse to its propagation direction $\hat{\vec{\kappa}}_l$. We use the notation $\hat{\vec{\kappa}}_l=\vec{\kappa}_l/|\vec{\kappa}_l|$ for the unit vector associated with the vector $\vec{\kappa}_l$. Correspondingly, we have $\vec{E}_l\cdot\vec{B}_l=0$ and $\hat{\vec{E}}_l\times\hat{\vec{B}}_l=\hat{\vec{\kappa}}_l$, as well as $\vec{E}_l={\cal E}_l\hat{\vec{E}}_l$ and $\vec{B}_l={\cal E}_l\hat{\vec{B}}_l$, with common field amplitude profile ${\cal E}_l$.
The total electric and magnetic fields associated with the collision of $l$ laser pulses follow upon superposition of the driving laser fields as $\vec{E}=\sum_l\vec{E}_l$ and $\vec{B}=\sum_l\vec{B}_l$, respectively.
For definiteness, here we consider only linearly polarized laser fields, such that the polarization vector of the pulse does not depend on the space-time coordinate and is given by the constant unit vector $\hat{\vec{E}}_l$. 

\subsection{Laser pulse profiles}\label{sec:profiles}

We model the pulse amplitude profile in position space as

\begin{equation}
	{\cal E}_l(x)={\mathfrak E}_{l}\,{\rm e}^{-\bigl(\frac{\hat{\vec{\kappa}}_l\cdot\vec{x}-t}{\tau_l/2}\bigr)^2}{\rm e}^{-\bigl(\frac{r_l}{w_l}\bigr)^2}\cos\bigl[\omega_l(\hat{\vec{\kappa}}_l\cdot\vec{x}-t)\big]\,, \label{eq:El}
\end{equation}
where we introduced the shorthand notation $r_l=\vec{x}-\hat{\vec{\kappa}}_l\cdot\vec{x}$.
Here, ${\mathfrak E}_l$ is the peak field amplitude, the first exponential function ensures a finite pulse duration $\tau_l$, and the second one a finite transverse extent characterized by the beam waist $w_l$. The pulse amplitude profile~\eqref{eq:El} squared can be expressed as

\begin{equation}
	{\cal E}_l^2(x)={\mathfrak E}_{l}^2\,{\rm e}^{-2\bigl(\frac{\hat{\vec{\kappa}}_l\cdot\vec{x}-t}{\tau_l/2}\bigr)^2}{\rm e}^{-2\bigl(\frac{r_l}{w_l}\bigr)^2}\,\frac{1}{2}\Bigl(1+\cos\bigl[2\omega_l(\hat{\vec{\kappa}}_l\cdot\vec{x}-t)\big]\Bigr)\,.\label{eq:El2}
\end{equation}
For completeness, note that the expression in Eq.~\eqref{eq:El} follows from the leading order paraxial Gaussian beam solution of the wave equation \cite{Siegman} supplemented with a Gaussian temporal envelope \cite{Karbstein:2015cpa} upon formally sending the Rayleigh range to infinity, while keeping the beam waist fixed \cite{Gies:2017ygp}.
Also note that in general the above laser pulse model tends to somewhat overestimate signatures of quantum vacuum nonlinearity: it does not account for the widening of the transverse beam profile as a function of the longitudinal coordinate with increasing distance from the focal spot, coming along with the reduction of the on-axis peak field amplitude.

As demonstrated in Ref.~\cite{Gies:2017ygp}, the approximation~\eqref{eq:El} nevertheless allows for reasonable estimates of all-optical signatures of QED vacuum nonlinearity in laser pulse collisions, where the signal photons are predominantly induced in the space-time region where the strong electromagnetic fields of the colliding pulses overlap.
Though the longitudinal decay of the individual laser fields is not explicitly accounted for, in this case the finite pulse durations of the colliding pulses naturally confine the strong-field region sourcing the signal.
Substantial deviations between the approximative and full results are only to be expected for the special cases of a single driving laser pulse, or almost copropagating driving pulses.
The latter situations are, however, not of phenomenological interest as their signal photon yield is completely negligible if not zero.
Obviously, no signal photons are induced in the presence of a single paraxial laser pulse as introduced above:
due to the fact that the electric and magnetic fields characterizing the pulse are always orthogonal to each other and equal in amplitude, we have ${\cal F}={\cal G}=0$, and the signal photon current~\eqref{eq:jk} vanishes identically in this case. 

The laser pulse energy follows upon integration of Eq.~\eqref{eq:El2} over the spatial coordinates; cf. Eq.~\eqref{eq:W}. This yields

\begin{equation}
 W_l=\int{\rm d}^3x\,{\cal E}_l^2(x)
 \approx\frac{1}{4}\Bigl(\frac{\pi}{2}\Bigl)^{\frac{3}{2}}\tau_lw_l^2 {\mathfrak E}_{l}^2\,,\label{eq:Wl}
\end{equation}
where we neglected subleading terms of ${\cal O}(\frac{1}{\tau_l\omega_l})$; see Ref.~\cite{Karbstein:2017jgh} for a detailed discussion. Obviously, the dimensionless quantity $\frac{1}{\tau_l\omega_l}$ corresponds to a small parameter if the pulse consists of many cycles, as is typically the case for currently available high-intensity laser pulses, justifying this approximation. Equation~\eqref{eq:Wl} allows us to express the peak field amplitude in terms of the pulse energy, pulse duration and beam waist as

\begin{equation}
	{\mathfrak E}_{l}\simeq 2\Bigl(\frac{8}{\pi}\Bigr)^{\frac{1}{4}}\sqrt{\frac{W_l}{\tau_l w_l^2\pi}}\,. \label{eq:E0l}
\end{equation}
Apart from an overall numerical factor, the peak field amplitude~\eqref{eq:E0l} is proportional to the square root of the pulse energy divided by the product of the pulse duration and the transverse focus area.  

Subsequently, we consider two exemplary collision scenarios of FEL and high-intensity laser pulses, and explicitly determine the attainable numbers of signal photons encoding the signature of quantum vacuum nonlinearity in the prescribed electromagnetic fields. To this end, we resort to the above approximations. One of our main interests is on the perspectives of measuring the respective signal photons with state-of-the-art technology.
As detailed above, in order to obtain a measurable response, we need at least two colliding laser pulses.

\subsection{Head-on collision of two laser pulses}\label{sec:headon}

\begin{figure}
\centering
 \includegraphics[width=0.7\linewidth]{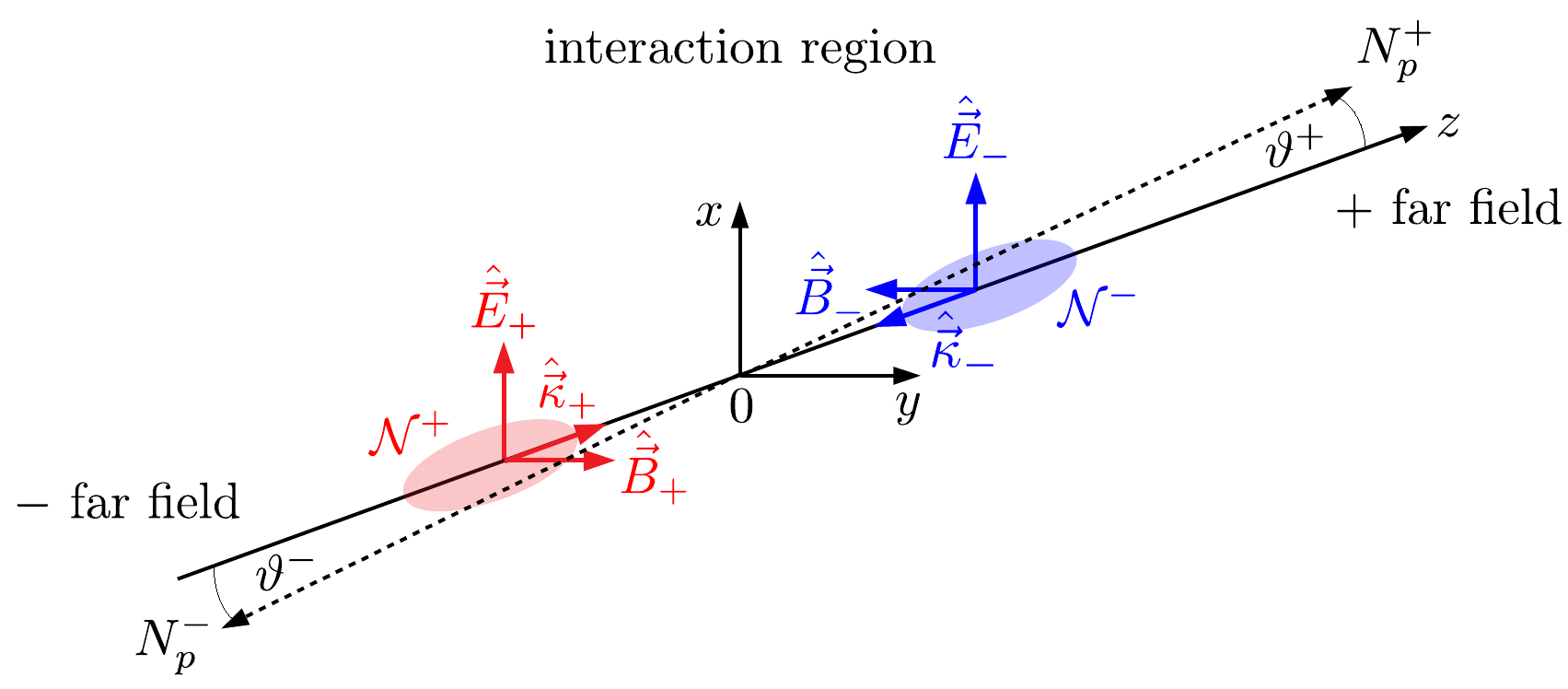}
 \caption{Illustration of the head-on collision geometry considered in Sec.~\ref{sec:headon} for $\phi_\pm=0$. The two driving $\pm$ laser pulses comprising ${\cal N}_\pm$ photons collide at $\vec{x}=0$.
 The quasielastically scattered signal photons $N^\pm_p$ encoding the signature of quantum vacuum nonlinearity are detected in the $\pm$ far field.}
 \label{fig:Headon}
\end{figure}

In the next step, we study optical signatures of quantum vacuum nonlinearity in the head-on collision of two laser pulses, which we label by $l=\pm$.
For simplicity, we only focus on optimal collisions at zero impact parameter.
Without loss of generality, the $+$ ($-$) pulse is assumed to propagate in positive (negative) $z$ direction, i.e., $\hat{\vec{\kappa}}_\pm=\pm\hat{\vec{e}}_z$. Here, $\hat{\vec{e}}_z$ is the unit vector pointing in $z$ direction.  
In turn, the associated electric and magnetic field vectors can be parameterized as
$\vec{E}_\pm={\cal E}_\pm(x)\hat{\vec{E}}_\pm$ and $\vec{B}_\pm={\cal E}_\pm(x)\hat{\vec{B}}_\pm$, with

\begin{equation}
 \hat{\vec{E}}_\pm=\left(\begin{array}{c}
                            \cos\phi_\pm \\ \sin\phi_\pm \\ 0
                           \end{array}
\right)\quad \text{and}\quad
 \hat{\vec{B}}_\pm=\pm\left(\begin{array}{c}
                            -\sin\phi_\pm \\ \cos\phi_\pm \\ 0
                           \end{array}
\right)\,.
\end{equation}
The angle parameters $\phi_\pm$ allow for the parameterization of all the possible polarization configurations of the two head-on colliding, linearly polarized laser pulses.
See Fig.~\ref{fig:Headon} for an illustration of the collision geometry highlighting various relevant parameters.
Sticking to these definitions, the polarization and magnetization vectors~\eqref{eq:PMleading} become

\begin{align}
 \vec{P}&\simeq\sqrt{\frac{\alpha}{\pi}}\frac{m_e^2}{45\pi}\Bigl(\frac{e}{m_e^2}\Bigr)^3\sum_{l=\pm}\bigl[4\hat{\vec{E}}_l\cos(\phi_+-\phi_-)-7\hat{\vec{B}}_l\sin(\phi_+-\phi_-)\bigr]{\cal E}_+(x){\cal E}_-(x){\cal E}_l(x)\,, \nonumber\\
 \vec{M}&\simeq\sqrt{\frac{\alpha}{\pi}}\frac{m_e^2}{45\pi}\Bigl(\frac{e}{m_e^2}\Bigr)^3\sum_{l=\pm}\bigl[4\hat{\vec{B}}_l\cos(\phi_+-\phi_-)+7\hat{\vec{E}}_l\sin(\phi_+-\phi_-)\bigr]{\cal E}_+(x){\cal E}_-(x){\cal E}_l(x)\,, \label{eq:PMleading+-}
\end{align}
and the amplitude~\eqref{eq:epj}, whose modulus square results in the differential signal photon number~\eqref{eq:dNp}, can be schematically written as

\begin{equation}
 \vec{\epsilon}_p(k)\cdot\vec{j}(k)={\rm i}c_+\int{\rm d}^4x\,{\rm e}^{-{\rm i}kx} {\cal E}_+(x){\cal E}_-^2(x)
 +{\rm i}c_-\int{\rm d}^4x\,{\rm e}^{-{\rm i}kx} {\cal E}_-(x){\cal E}_+^2(x)\,, \label{eq:epj+-}
\end{equation}
with real-valued coefficients $c_\pm$ containing information about the alignment of the driving fields and the polarization properties of the induced signal photons.
Hence, the explicit determination of the differential signal photon number essentially reduces to performing the Fourier transform of the monomials ${\cal E}_+(x){\cal E}_-^2(x)$ and ${\cal E}_+^2(x){\cal E}_-(x)$.

To further simplify the following considerations, we assume the colliding pulses to feature the same duration, $\tau_+=\tau_-=\tau$, and to be focused to the same waist, $w_+=w_-=w$.
Moreover, we explicitly limit ourselves to quasielastically scattered signal photon contributions only. 
The latter naturally decompose into two distinct contributions, namely those made up of laser photons of the $+$ pulse scattered at the intensity profile, or equivalently field amplitude squared, of the $-$ pulse, and vice versa.
These contributions can clearly be associated with the terms proportional to ${\cal E}_+(x){\cal E}_-^2(x)$ and ${\cal E}_-(x){\cal E}_+^2(x)$ in Eq.~\eqref{eq:PMleading+-}.
Quasielastically scattered signal photons emerging from the $\pm$ pulse are characterized by a photon energy very similar to the one of the laser photons constituting the $\pm$ pulse, and are predominantly scattered in its forward direction.
They in particular do not depend on the oscillation frequency ($\simeq$ photon energy) of the laser pulse acting as scatterer.
Correspondingly, the cosine term in the pulse amplitude squared \eqref{eq:El2} with $l=\mp$ can be associated with manifestly inelastic signal photon contributions, and thus be neglected when performing the Fourier transform of

\begin{equation}
 {\cal E}_\pm(x){\cal E}_\mp^2(x)\to\frac{1}{4}{\mathfrak E}_{\pm}{\mathfrak E}_{\mp}^2\,{\rm e}^{-\frac{4}{\tau^2}(3z^2+3t^2\pm2zt)-3\frac{x^2+y^2}{w^2}}\sum_{l=\pm}{\rm e}^{{\rm i}l\omega_\pm(z\mp t)}\,. \label{eq:EpmEmp2}
\end{equation}
The Fourier transform of Eq.~\eqref{eq:EpmEmp2} entering Eq.~\eqref{eq:epj} can be performed straightforwardly resulting in

\begin{equation}
 \int{\rm d}^4x\,{\rm e}^{-{\rm i}kx} {\cal E}_\pm(x){\cal E}_\mp^2(x)\to\frac{\sqrt{2}\pi^2}{192}{\mathfrak E}_{\pm}{\mathfrak E}_{\mp}^2 \tau^2w^2\,{\rm e}^{-\frac{w^2}{12}(k_x^2+k_y^2)}
 \sum_{l=\pm}{\rm e}^{-\frac{\tau^2}{16}[(\omega_\pm-l\frac{k_z\pm k^0}{2})^2+\frac{1}{2}(\frac{(k_z\mp k^0)}{2})^2]}\,. \label{eq:FT_EpmEmp2}
\end{equation}
Obviously, the exponential suppression in Eq.~\eqref{eq:FT_EpmEmp2} can be overcome for at least one of the summands if the conditions $k_x=k_y=0$, $k_z=\pm k^0$ and $k^0=\omega_\pm$ are fulfilled simultaneously.
These two sets of conditions govern the dominant signal photon emission channels, namely signal photons emitted in the forward cones of the driving lasers.
The complementary conditions with $k^0=-\omega_\mp$ will not play any role as they cannot be realized for manifestly positive $k^0$ and laser frequencies $\omega_\pm>0$.

One can easily convince oneself by an explicit calculation that the inclusion of the cosine term in Eq.~\eqref{eq:El2} results in manifestly inelastic contributions, characterized by the absorption/release of two laser photons associated with an energy transfer of $2\omega_\mp$ by the scatterer.

Due to the distinct directional emission properties of the dominant signal photon channels, interference terms between them arising upon taking the modulus square of Eq.~\eqref{eq:epj+-} receive an exponential suppression and can be safely neglected, such that

\begin{equation}
 \bigl|\vec{\epsilon}_p(k)\cdot\vec{j}(k)\bigr|^2\to c_+^2\,\Bigl|\int{\rm d}^4x\,{\rm e}^{-{\rm i}kx} {\cal E}_+(x){\cal E}_-^2(x)\Bigr|^2
 +c_-^2\,\Bigl|\int{\rm d}^4x\,{\rm e}^{-{\rm i}kx} {\cal E}_-(x){\cal E}_+^2(x)\Bigr|^2\,. \label{eq:epj+-quad}
\end{equation}
Therefore, the differential signal photon number~\eqref{eq:dNp} naturally decomposes into two contributions, ${\rm d}^3N_p\to {\rm d}^3N^+_p+{\rm d}^3N^-_p$, describing signal photons which are scattered quasielastically into the vicinity of the $+$ and $-$ direction, respectively.
Their differential numbers are given by

\begin{equation}
 {\rm d}^3N^\pm_p\simeq\frac{{\rm d}^3k}{(2\pi)^3}\,\frac{c_\pm^2}{2k^0}\,\Bigl|\int{\rm d}^4x\,{\rm e}^{-{\rm i}kx} {\cal E}_\pm(x){\cal E}_\mp^2(x)\Bigr|^2 \bigg|_{k^0=|\vec{k}|}\,. \label{eq:dNp+-}
\end{equation}

Subsequently, we aim at evaluating and analyzing these quantities.
For convenience, we make use of two slightly different parameterizations of the momentum coordinates, which are specifically tailored to the signal photon contributions ${\rm d}^3N^\pm_p$, namely $k_z=\pm{\rm k}\cos\vartheta_\pm$ and $k_x^2+k_y^2={\rm k}^2\sin^2\vartheta_\pm$, with ${\rm k}=|\vec{k}|=k^0$. The associated differential volume element is ${\rm d}^3k={\rm k}^2{\rm dk}\,{\rm d}\cos\vartheta_\pm\,{\rm d}\varphi$, where the angle $\varphi$ parameterizes rotations around the $z$ axis.
In these coordinates, the dominant signal photon contribution fulfills ${\rm k}\approx\omega_\pm$ and $\vartheta_\pm\approx0$.
Its differential number is expected to decline rapidly with increasing values of the polar angle $\vartheta_\pm$ and the energy difference $|{\rm k}-\omega_\pm|$; see Eq.~\eqref{eq:FT_EpmEmp2approx} below for an explicit confirmation of this expectation.

In the vicinity of $\vartheta_\pm=0$, we have $k_z\simeq\pm{\rm k}(1-\frac{\vartheta_\pm^2}{2})$, $k_x^2+k_y^2\simeq{\rm k}^2\vartheta_\pm^2$ and ${\rm d}^3k\simeq{\rm k}^2{\rm dk}\,\vartheta_\pm{\rm d}\vartheta_\pm\,{\rm d}\varphi$.
Keeping terms up to quadratic order in $\vartheta_\pm$ in the exponential of Eq.~\eqref{eq:FT_EpmEmp2} and neglecting the contribution in the sum over $l$ which is exponentially suppressed for ${\rm k}\approx\omega_\pm$, we obtain
\begin{equation}
 \int{\rm d}^4x\,{\rm e}^{-{\rm i}kx} {\cal E}_\pm(x){\cal E}_\mp^2(x)\to\frac{\sqrt{2}\pi^2}{192}{\mathfrak E}_{\pm}{\mathfrak E}_{\mp}^2 \tau^2w^2\,{\rm e}^{-\frac{w^2}{12}{\rm k}^2\vartheta_\pm^2}\,
 {\rm e}^{-\frac{\tau^2}{16}(\omega_\pm-{\rm k})^2-\frac{\tau^2}{32}{\rm k}(\omega_\pm-{\rm k})\vartheta_\pm^2}\,. \label{eq:FT_EpmEmp2approx}
\end{equation}
Moreover, in this limit the two transverse signal photon polarization vectors can be parameterized by a single angle variable as

\begin{equation}
 \vec{\epsilon}_{p}(k)=\left(\begin{array}{c}
                            \cos[\phi_\pm+\beta\pm\frac{\pi}{2}(p-1)] \\ \sin[\phi_\pm+\beta\pm\frac{\pi}{2}(p-1)] \\ 0
                           \end{array}
\right)\,. \label{eq:ep}
\end{equation}
As the orientation of the polarization basis defined below Eq.~\eqref{eq:dNp} is intimately linked to the direction of the signal photon's wave vector, also the polarization vectors~\eqref{eq:ep} depend on the emission direction. 
The choice of the angle $\beta$ fixes a specific linear polarization basis for the signal photons; without loss of generality our conventions are such that for $\beta=0$ we have $\vec{\epsilon}_{1}(k)=\hat{\vec{E}}_\pm$.
Making use of Eq.~\eqref{eq:ep} to explicitly work out the coefficients $c_\pm$, we obtain

\begin{equation}
  c_\pm\simeq k^0\sqrt{\frac{\alpha}{\pi}}\frac{m_e^2}{45\pi}\Bigl(\frac{e}{m_e^2}\Bigr)^3\bigl\{11\cos[\beta\pm\tfrac{\pi}{2}(p-1)]-3\cos[\beta\pm\tfrac{\pi}{2}(p-1)\pm2\Delta\phi]\bigr\}\,, \label{eq:cpm}
\end{equation}
where $\Delta\phi=\phi_+-\phi_-$ measures the angle between the polarization vectors of the two head-on colliding driving laser beams.
Plugging the results~\eqref{eq:FT_EpmEmp2approx} and \eqref{eq:cpm} into Eq.~\eqref{eq:dNp+-}, we find

\begin{align}
 {\rm d}^3N^\pm_p\simeq&\frac{{\rm d}^3k}{(2\pi)^3}\,\frac{1}{m_e^3}\frac{\tau{\rm k}}{(wm_e)^2}\sqrt{\frac{2}{\pi}}\,\Bigl(\frac{4\alpha^2}{135}\Bigr)^2\frac{W_\pm}{m_e} \Bigl(\frac{W_\mp}{m_e}\Bigr)^2\,{\rm e}^{-\frac{w^2}{6}{\rm k}^2\vartheta_\pm^2}\,
 {\rm e}^{-\frac{\tau^2}{8}(\omega_\pm-{\rm k})^2-\frac{\tau^2}{16}{\rm k}(\omega_\pm-{\rm k})\vartheta_\pm^2}
  \nonumber\\
 &\times\bigl\{11\cos[\beta\pm\tfrac{\pi}{2}(p-1)]-3\cos[\beta\pm\tfrac{\pi}{2}(p-1)\pm2\Delta\phi]\bigr\}^2\,, \label{eq:d3Napp}
\end{align}
where we expressed the peak field amplitudes of the driving laser fields by the respective pulse energies $W_\pm$ via Eq.~\eqref{eq:E0l}.

Due to the fact that for $\vartheta_\pm\ll1$ the main contribution to the integral over $\rm k$ stems from $k\simeq\omega_\pm$, it should amount to a good approximation to set ${\rm k}=\omega_\pm$ in the terms in the exponential which are proportional to $\vartheta_\pm^2$, as well as in the overall monomial of $\rm k$ multiplying the exponential function. Formally extending the integration over $\rm k$ as $\int_0^\infty{\rm dk}\to\int_{-\infty}^\infty{\rm dk}$, the integral over $\rm k$ is then promoted to a simple Gaussian integral. Implementing this approximation in Eq.~\eqref{eq:d3Napp}, upon performing the trivial integration over the azimuthal angle $\varphi$ we finally get the following concise expression for the differential number of induced signal photons

\begin{align}
 {\rm d}^2N^\pm_p\simeq&\,{\rm d}\vartheta_\pm \vartheta_\pm{\rm dk}\,\tau\sqrt{\frac{2}{\pi}}\,\Bigl(\frac{2\alpha^2}{135\pi}\Bigr)^2\frac{1}{(w m_e)^2}\Bigl(\frac{\omega_\pm}{m_e}\Bigr)^3\frac{W_\pm}{m_e} \Bigl(\frac{W_\mp}{m_e}\Bigr)^2\,
 {\rm e}^{-\frac{\tau^2}{8}(\omega_\pm-{\rm k})^2}\,{\rm e}^{-\frac{w^2}{6}\omega_\pm^2\vartheta_\pm^2}  \nonumber\\
 &\times\bigl\{11\cos[\beta\pm\tfrac{\pi}{2}(p-1)]-3\cos[\beta\pm\tfrac{\pi}{2}(p-1)\pm2\Delta\phi]\bigr\}^2\,. \label{eq:d3Napprox}
\end{align}
This result encodes the distribution of the signal photons as a function of their energy $\rm k$ and scattering angle $\vartheta_\pm$.
The fact that the signal photons do not exhibit any nontrivial dependence on the angle $\varphi$ is rooted in the rotational symmetry of the considered head-on collision geometry around the ${\rm z}$ axis.

The differential number of signal photons attainable in a polarization insensitive measurement follows from Eq.~\eqref{eq:d3Napprox} upon summation over the two transverse modes as ${\rm d}^2N^\pm=\sum_{p=1}^2 {\rm d}^2N_p^\pm$.
On the other hand, the differential number of signal photons scattered into a perpendicularly polarized mode ${\rm d}^2N^\pm_\perp$ is obtained upon setting $p=1$ and $\beta=\frac{\pi}{2}$.
Note, that these polarization-flipped signal photons are conventionally associated with the non-linear QED phenomenon of vacuum birefringence \cite{Toll:1952,Baier,BialynickaBirula:1970vy,Aleksandrov:1985,Kotkin:1996nf}: an electromagnetic field can effectively endow the quantum vacuum with two different indices of refraction, associated with photon polarization eigenmodes in this field configuration. Hence, when traversing a strong electromagnetic field region, originally linearly polarized probe photons with polarization overlap to both of these eigenmodes can pick up a tiny ellipticity, thereby effectively populating a perpendicularly polarized mode \cite{Dinu:2013gaa,Karbstein:2015xra}. The number of polarization-flipped signal photons constitutes the signature of vacuum birefringence in proposals to measure the effect of vacuum birefringence in high-intensity laser experiments \cite{Heinzl:2006xc}; cf. also Refs.~\cite{DiPiazza:2006pr,Dinu:2013gaa,Dinu:2014tsa,Karbstein:2015xra,Schlenvoigt:2016,Karbstein:2016lby,King:2016jnl,Bragin:2017yau,Karbstein:2017jgh,Karbstein:2018omb}. See Ref.~\cite{Karbstein:2016hlj} for a pedagogical presentation of the conventional derivation of vacuum birefringence in constant background fields, and Ref.~\cite{Battesti:2018bgc} for a concise review of experimental activities aiming at measuring vacuum birefringence in quasi-constant magnetic fields, actively pursued by various experiments \cite{Cantatore:2008zz,Berceau:2011zz,Fan:2017fnd}.

Upon performing the integration over the signal photon energy $\rm k$, the differential number of signal photons attainable in a polarization insensitive measurement as well as the differential number of polarization-flipped signal photons can be compactly expressed as

\begin{align}
\left\{\!\!\!
\begin{array}{c}
 {\rm d}N^\pm \\ {\rm d}N^\pm_\perp
\end{array}
\!\!\!\right\}
 \simeq{\rm d}\vartheta_\pm \vartheta_\pm\,\Bigl(\frac{4\alpha^2}{135\pi}\Bigr)^2\frac{1}{(w m_e)^2}\Bigl(\frac{\omega_\pm}{m_e}\Bigr)^3\frac{W_\pm}{m_e} \Bigl(\frac{W_\mp}{m_e}\Bigr)^2
\,{\rm e}^{-\frac{1}{6}(\omega_\pm w)^2\vartheta_\pm^2}
\left\{\!\!\!
\begin{array}{c}
 130-66\cos(2\Delta\phi) \\ 9\sin^2(2\Delta\phi)
\end{array}
\!\!\!\right\}\,. \label{eq:dNtotperp}
\end{align}
Interestingly, upon integration over the signal photon energy, the dependence of the pulse duration $\tau$ drops out completely within the considered approximation, such that Eq.~\eqref{eq:dNtotperp} becomes $\tau$ independent.
Moreover, note that this expression implies that the signal attainable in a polarization insensitive measurement is maximized for $\Delta\phi=\frac{\pi}{2}$, while the signal scattered into a perpendicularly polarized mode is maximized for $\Delta\phi=\frac{\pi}{4}$ \cite{Karbstein:2019bhp}.

From Eq.~\eqref{eq:dNtotperp}, we can straightforwardly read of the angular divergence of the scattering signal with respect to the polar angle $\vartheta_\pm$.
The $1/{\rm e}^2$ radial beam divergence of the scattering signal in $\pm$ direction is $\theta_{{\rm signal},\pm}\simeq\frac{2\sqrt{3}}{w\omega_\pm}=\frac{\sqrt{3}\lambda_\pm}{\pi w}$.
This value is to be compared with the far-field radial beam divergences of the driving laser beams focused to a spot size of radius $w$, which we assume to be well-described by the paraxial Gaussian beam result $\theta_\pm\simeq\frac{\lambda_\pm}{\pi w}$ \cite{Siegman}.
Hence, we have $\theta_{{\rm signal},\pm}\simeq\sqrt{3}\,\theta_\pm$, implying that the divergence of the signal in general surpasses the divergence of the respective driving laser pulse.

The far-field angular distribution of the photons constituting the $\pm$ laser pulse should be well-described by

\begin{equation}
 {\rm d}{\cal N}^\pm\simeq{\rm d}\vartheta_\pm \vartheta_\pm\, {\cal N}^\pm\,(\omega_\pm w)^2
 \,{\rm e}^{-\frac{1}{2}(\omega_\pm w)^2\vartheta_\pm^2}\,, \label{eq:d2N}
\end{equation}
where, ${\cal N}^\pm\simeq W_\pm/\omega_\pm$ denotes the number of laser photons constituting the $\pm$ pulse.
Note, that the different angular decay of the driving laser photons~\eqref{eq:d2N} and the signal photons~\eqref{eq:dNtotperp} with $\vartheta_\pm$ ensures that even if the signal is completely background dominated for $\vartheta_\pm=0$, as is typically the case, it will eventually surpass the background of the driving laser photons from a certain finite value of $\vartheta_\pm$ onwards. However, the important question to be answered in this context is if the number of signal photons which can be distinguished from the background is large enough to allow for a measurement of the effect for given laser parameters. In the following we elaborate on this question on a more quantitative level.

To this end, we assume that the differential number of signal photons attainable in a polarization insensitive measurement fulfills ${\rm d}N^\pm(\vartheta_\pm=0)>{\cal P}^\pm{\rm d}{\cal N}^\pm(\vartheta_\pm=0)$, were the constant ${\cal P}^\pm$ quantifies the sensitivity of the considered experiment. In turn, only those signal photons fulfilling the discernibility criterion

\begin{equation}
 {\rm d}N^\pm(\vartheta_\pm)\geq {\cal P}^\pm{\rm d}{\cal N}^\pm(\vartheta_\pm)\,, \label{eq:dNPdN}
\end{equation}
can in principle be measured in experiment.
Assuming that the polarization-flipped signal photons similarly fulfill ${\rm d}N_\perp^\pm(\vartheta_\pm=0)>{\cal P}_\perp^\pm{\rm d}{\cal N}^\pm(\vartheta_\pm=0)$, the analogous discernibility criterion for the polarization-flipped signal photons reads

\begin{equation}
 {\rm d}N_\perp^\pm(\vartheta_\pm)\geq {\cal P}_\perp^\pm{\rm d}{\cal N}^\pm(\vartheta_\pm)\,. \label{eq:dNperpPdN}
\end{equation}
Some comments are in order here, to clarify that even though the criteria~\eqref{eq:dNPdN} and \eqref{eq:dNperpPdN} are formally very similar, the values of ${\cal P}$ and ${\cal P}_\perp$ attainable in experiment are typically very different.
It is easily conceivable that ${\cal P}_\perp$ can be a very small number in experiment, where it characterizes the achievable quality or {\it purity} of polarization filtering.
While an ideal polarization filter would even achieve ${\cal P}_\perp=0$, any realistic polarization filter inevitably fulfills ${\cal P}_\perp\neq0$.
In particular in the x-ray regime, polarization purities ${\cal P}_\perp$ on the level of $10^{-10}$ have been demonstrated \cite{Marx:2011,Marx:2013xwa,Schulze:2018}. 
On the other hand, it seems reasonable to assume that the analogous quantity $\cal P$ characterizing the sensitivity to be able to detect signal photons which are -- apart from their wider scattering with $\vartheta_\pm$ -- indistinguishable from the driving laser photons, can at best reach an value of ${\cal P}\geq1$, independent of the laser frequencies.

Equations~\eqref{eq:dNPdN} and \eqref{eq:dNperpPdN} can be staightforwardly solved for the angles for which the signal photons can be discerned from the background of the driving laser photons.
The latter are constrained by

\begin{align}
 \vartheta_\pm^2 \gtrsim\,\frac{3}{(\omega_\pm w)^2}\Biggl[2\ln\Bigl(\frac{135\pi}{4\alpha^2}(wm_e)^2\frac{m_e}{\omega_\pm}\frac{m_e}{W_\mp}\Bigr)-
\left\{\!\!\!
\begin{array}{c}
 \ln\bigl[130-66\cos(2\Delta\phi)\bigr]-\ln{\cal P}^\pm \\
 \ln\bigl[9\sin^2(2\Delta\phi)\bigr]-\ln{\cal P}^\pm_\perp
\end{array}
\!\!\!\right\}\Biggr]\,, \label{eq:theta=}
\end{align}
where the upper line is the result for a polarization insensitive measurement, and the lower line the one for the polarization-flipped signal.

Subsequently, we denote the signal photons fulfilling the above discernibility criteria~\eqref{eq:dNPdN} and \eqref{eq:dNperpPdN} by $\tilde N^\pm$ and $\tilde N^\pm_\perp$, respectively.
To determine their numbers, we integrate Eq.~\eqref{eq:dNtotperp} over all values of $\vartheta_\pm$ fulfilling Eq.~\eqref{eq:theta=}. To this end, we formally extend the integration over $\vartheta_\pm$ from the threshold value up to infinity. This resuls in

\begin{align}
 \left\{\!\!\!
\begin{array}{c}
 \tilde N^\pm \\ \tilde N_\perp^\pm
\end{array}
\!\!\!\right\}
 \simeq\,\Bigl(\frac{4\alpha^2}{135\pi}\Bigr)^3\frac{3}{(w m_e)^6} \Bigl(\frac{\omega_\pm}{m_e}\Bigr)^2 \frac{W_\pm}{m_e} \Bigl(\frac{W_\mp}{m_e}\Bigr)^3
\left\{\!\!\!
\begin{array}{c}
 \bigl[130-66\cos(2\Delta\phi)\bigr]^{3/2}/\sqrt{{\cal P}^\pm} \\ \bigl[9\sin^2(2\Delta\phi)\bigr]^{3/2}/\sqrt{{\cal P}^\pm_\perp}
\end{array}
\!\!\!\right\}\,. \label{eq:Ntilde}
\end{align}
Finally, for comparison we also evaluate the respective total numbers of signal photons. Integrating Eq.~\eqref{eq:dNtotperp} over all values of $0\leq\vartheta_\pm\leq\infty$, yields

\begin{align}
\left\{\!\!\!
\begin{array}{c}
 N^\pm \\ N^\pm_\perp
\end{array}
\!\!\!\right\}
 \simeq\Bigl(\frac{4\alpha^2}{135\pi}\Bigr)^2\frac{3}{(w m_e)^4}\frac{\omega_\pm}{m_e}\frac{W_\pm}{m_e} \Bigl(\frac{W_\mp}{m_e}\Bigr)^2
\left\{\!\!\!
\begin{array}{c}
 130-66\cos(2\Delta\phi) \\ 9\sin^2(2\Delta\phi)
\end{array}
\!\!\!\right\}\,. \label{eq:Ntotperp}
\end{align}

To obtain a feeling for the size of the effect in prospective experiments based on state-of-the-art technology, we subsequently focus on two different experimental scenarios: the collision of two near-infrared laser pulses delivered by petawatt-class high-intensity laser systems, and the collision of such a high-intensity laser pulse with an intense x-ray pulse provided by an FEL. The first scenario is possible at various high-field laboratories worldwide, for instance ELI \cite{ELI}, among many others. The second one requires an high-intensity laser to be installed at an FEL, such as at the Helmholtz International Beamline for Extreme Fields (HIBEF) \cite{HIBEF} at the European X-Ray Free-Electron Laser (XFEL) facility \cite{XFEL}. For recent activities at the Spring-8 Angstrom Compact Free Electron Laser (SACLA) facility \cite{SACLA}, see Refs.~\cite{Inada:2017lop,Seino:2019wkb}.
In the remainder of these notes, we exclusively stick to the polarization alignments maximizing the signal, i.e., adopt the choice of $\Delta\phi=\frac{\pi}{2}$ for $N^\pm$, and $\Delta\phi=\frac{\pi}{4}$ for $N^\pm_\perp$, respectively.

\subsubsection{Head-on collision of high-intensity laser pulses}\label{sec:bothHIL}

For definiteness, we assume two identical high-intensity lasers of wavelength $\lambda_\pm=\lambda$ and energy $W_\pm=W$ at our disposal. Both pulses are assume to be optimally focussed to their diffraction limit, corresponding to the minimum possible beam waist $w=\lambda$.
With these assumptions Eqs.~\eqref{eq:Ntilde} and \eqref{eq:Ntotperp} can be expressed as

\begin{align}
\left\{\!\!\!
\begin{array}{c}
 N^\pm \\ N^\pm_\perp
\end{array}
\!\!\!\right\}
 \simeq\Bigl(\frac{4\alpha^2}{135}\Bigr)^2 \frac{6}{\pi}\frac{1}{(\lambda m_e)^5}\Bigl(\frac{W}{m_e}\Bigr)^3
\left\{\!\!\!
\begin{array}{c}
 196 \\ 9
\end{array}
\!\!\!\right\}
\approx
\left\{\!\!\!
\begin{array}{c}
 1.46\times10^{-2} \\ 6.68\times10^{-4}
\end{array}
\!\!\!\right\}\Bigl(\frac{W}{1\,{\rm J}}\Bigr)^3\Bigl(\frac{1\,\si\micro{\rm m}}{\lambda}\Bigr)^5
\label{eq:N_twoHILs}
\end{align}
and

\begin{align}
 \left\{\!\!\!
\begin{array}{c}
 \tilde N^\pm \\ \tilde N_\perp^\pm
\end{array}
\!\!\!\right\}
 \simeq\Bigl(\frac{4\alpha^2}{135}\Bigr)^3 \frac{6}{\pi}\frac{1}{(\lambda m_e)^8}\Bigl(\frac{W}{m_e}\Bigr)^4
\left\{\!\!\!
\begin{array}{c}
 5488/\sqrt{{\cal P}^\pm} \\ 54/\sqrt{{\cal P}^\pm_\perp}
\end{array}
\!\!\!\right\}
\approx
\left\{\!\!\!
\begin{array}{c}
 2.82\times10^{-14}/\sqrt{{\cal P}^\pm} \\ 2.79\times10^{-16}/\sqrt{{\cal P}^\pm_\perp}
\end{array}
\!\!\!\right\}\Bigl(\frac{W}{1\,{\rm J}}\Bigr)^4\Bigl(\frac{1\,\si\micro{\rm m}}{\lambda}\Bigr)^8\,
\label{eq:disN_twoHILs}
\end{align}
respectively.

For the two identical $10$ Petawatt (PW) lasers available at ELI-NP \cite{ELI}, delivering pulses of energy $W=200\,{\rm J}$ and duration $\tau=20\,{\rm fs}$ at a wavelength of $\lambda=800\,{\rm nm}$ and a repetition rate of one shot per minute, Eq.~\eqref{eq:N_twoHILs} results in a total number of $N^\pm\approx355584$ ($N_\perp^\pm\approx16384$) signal photons per shot.
Assuming two $1\,{\rm PW}$ lasers, delivering pulses of a tenth of this energy but otherwise equivalent parameters at our disposal, these numbers are reduced by a factor of $10^{-3}$. 
While these values might seem quite promising, one should always keep in mind that the relevant quantity deciding if a QED vacuum signature is accessible in experiment is not the total number of induced signal photons, but the number of signal photons which are discernible from the background of the driving laser beams.
Their explicit numbers follow from Eq.~\eqref{eq:disN_twoHILs}. Even for an optimal sensitivity of ${\cal P}=1$, the number of discernible signal photons per shot attainable in a polarization insensitive measurement is as small as $\tilde N^\pm\approx2.69\times10^{-4}$. On the other hand, in order to be able to detect at least one polarization-flipped signal photon in $\pm$ direction per shot, the polarization purity should be as good as ${\cal P}_\perp^\pm\approx7.08\times10^{-12}$.
For near-infrared frequencies and the considered large laser-photon fluxes, this value is out of reach at the moment, which is why we do not even bother to provide the explicit values of the threshold angles~\eqref{eq:theta=} delimiting the discernible signal regime from below.

From this analysis, we conclude that though the head-on collision of two high-intensity laser pulses of near-infrared frequencies as delivered by state-of-the-art high-intensity lasers of the Petawatt class gives rise to a number of quasielastically scattered signal photons per shot, in experiment these signal photons are essentially indiscernible from the driving laser photons. Besides, it should be noted that the inevitable presence of shot-to-shot fluctuations resulting in collisions with non-optimal impact parameter in experiment tends to further diminish the effect \cite{King:2012aw,Dinu:2014tsa,Karbstein:2018omb}.

For completeness, let us emphasize here that it is nevertheless possible to come up with laser-pulse collision scenarios exclusively based on near-infrared high-intensity lasers and state-of-the-art detection technology, which give rise to several discernible signal photons per shot \cite{Moulin:2002ya,Lundstrom:2005za,Gies:2017ezf,King:2018wtn,Karbstein:2019dxo}.
These scenarios typically involve the collision of more than two focused high-intensity laser pulses at finite angles and heavily rely on the availability of different-color driving beams, as, e.g., accessible by sum-frequency generation techniques.
It is intuitively clear that the many additional degrees of freedom becoming available when considering collision scenarios involving multiple beams of different colors should allow for the tailoring of optimal emission signals.
At best, the latter give rise to signal photons which are characterized by distinct frequencies not contained in the spectrum of the driving beams, and are emitted into directions outside the forward cones of the driving laser beams, thereby allowing for an excellent signal-to-background separation.

It should be noted that the study detailed here for the special case of two head-on colliding laser pulses can be readily extended to such scenarios.
However, particularly due to the large number of parameters characterizing the latter case, the explicit calculations needed for its theoretical analysis in general become much more tedious.

The most immediate and logical extension of the presented matters rather seems to be a careful re-analysis of the special scenario investigated here, but this time explicitly accounting for inelastic signal photon contributions.
While these contributions are certainly characterized by photon energies not contained in the spectra of the driving laser pulses, the associated photon numbers are generically much smaller than the numbers of quasielastically scattered signal photons \cite{Gies:2017ygp,Karbstein:2014fva}.
On the other hand, as emphasized in these notes, the total numbers of signal photons contain only limited relevant information. The important quantity to be analyzed instead is the number of discernible signal photons. It contains the information if a nonlinear QED signature is actually accessible in experiment. In fact, given the extremely tiny values found in Eq.~\eqref{eq:disN_twoHILs} for the numbers of discernible signal photons in the quasielastic channel, it is plausible that -- though very probably still not measurable with state-of-the-art technology -- the full discernible signal will be dominated by inelastic contributions; cf. also the recent study~\cite{Karbstein:2019dxo}.

\subsubsection{Head-on collision of high-intensity and free-electron laser pulses}\label{sec:biref}

In a next step, we focus on the laser pulse collision scenario involving both high-intensity laser and FEL pulses.
Our conventions are such that the FEL (high-intensity laser) pulse propagates in $+$ ($-$) direction. 
To be specific, we assume the x-ray pulse to comprise ${\cal N}^+=10^{12}$ photons at an energy of $\omega_+=12914\,{\rm eV}$, corresponding to a pulse energy of $W_+\simeq{\cal N}^+\omega_+\approx 2.07\times10^{-3}\,{\rm J}$.
For photons of this particular energy a polarization purity of ${\cal P}_\perp=5.7\times10^{-10}$ has been demonstrated in experiment \cite{Marx:2013xwa}.
Due to the rather small value of $W_+$ available in the present scenario and taking into account that $w\gtrsim\lambda_-$, a comparison with our findings in Sec.~\ref{sec:bothHIL} immediately implies that the signal photon contributions induced in $-$ direction are very tiny and not accessible in experiment with current technology.
Correspondingly, we neglect them from the outset, and exclusively focus on the $+$ channel. The latter encompasses signal photons of x-ray energy, which are scattered quasielastically at the high-intensity laser pulse.
Aiming at the determination of their prospective numbers attainable in experiment, we assume the detection sensitivities to be given by ${\cal P}^+=1$ and ${\cal P}_\perp^+=5.7\times10^{-10}$, respectively.

Plugging these parameters into Eqs.~\eqref{eq:Ntilde} and \eqref{eq:Ntotperp}, we obtain

\begin{align}
\left\{\!\!\!
\begin{array}{c}
 N^+ \\ N^+_\perp
\end{array}
\!\!\!\right\}
 \approx
 \left\{\!\!\!
\begin{array}{c}
 3.15\times10^{-1} \\ 1.44\times10^{-2}
\end{array}
\!\!\!\right\}\Bigl(\frac{W_-}{1\,{\rm J}}\Bigr)^2\Bigl(\frac{1\,\si\micro{\rm m}}{w}\Bigr)^4 \label{eq:Nx}
\end{align}
for the total numbers of induced signal photons, and 

\begin{align}
 \left\{\!\!\!
\begin{array}{c}
 \tilde N^+ \\ \tilde N_\perp^+
\end{array}
\!\!\!\right\}
 \approx
 \left\{\!\!\!
\begin{array}{c}
 6.35\times10^{-9} \\ 2.61\times10^{-6}
\end{array}
\!\!\!\right\}\Bigl(\frac{W_-}{1\,{\rm J}}\Bigr)^3\Bigl(\frac{1\,\si\micro{\rm m}}{w}\Bigr)^6 \label{eq:Ndisx}
\end{align}
for the corresponding discernible signal photon numbers.
The angular regime~\eqref{eq:theta=} into which the latter are emitted is constrained by 
\begin{align}
 \vartheta_+ \gtrsim\,37.41\frac{1\,\si\micro{\rm m}}{w}\sqrt{
c_0-\ln\Bigl[\frac{W_-}{1\,{\rm J}}\Bigl(\frac{1\,\si\micro{\rm m}}{w}\Bigr)^2\Bigr]}\,\si\micro{\rm rad}\,,
\end{align}
with $c_0\simeq14.94$ for the case of a polarization insensitive measurement, and $c_0\simeq5.84$ for the polarization-flipped signal.

Subsequently, we adopt the parameters of the high-intensity laser installed at the HIBEF beamline \cite{HIBEF} at the European XFEL \cite{XFEL}. This should serve as an explicit example to illustrate the prospective signal photon numbers attainable at a facility currently in operation. The high-intensity laser at HIBEF is designed to deliver pulses of energy $W_-=10\,{\rm J}$ and duration $\tau=25\,{\rm fs}$ at a wavelength of $\lambda_-=800\,{\rm nm}$ and a repetition rate of $5\,{\rm Hz}$.
In order to maximize the signal, we consider the beam waist $w$ to be given by the minimal achievable value.
This is achieved by focusing the high-intensity laser beam to its diffraction limit, implying that $w=\lambda_-$.

Plugging these parameters into Eqs.~\eqref{eq:Nx} and \eqref{eq:Ndisx}, we arrive at a total number of $N^+\approx77$ ($N^+_\perp\approx4$) signal photons per shot. The numbers of discernible signal photons per shot are $\tilde N^+\approx2.42\times10^{-5}$ and $\tilde N^+_\perp\approx 9.96\times10^{-3}$, to be detected at polar angles fulfilling $\vartheta_+\gtrsim163\,\si\micro{\rm rad}$ and $\vartheta_+\gtrsim82\,\si\micro{\rm rad}$, respectively.
Taking into account the repetition rate of the high-intensity laser, these results imply $\tilde N\approx0.44$ and $\tilde N_\perp\approx179$ discernible signal photons per hour.
In particular the last number highlights that a measurement of QED vacuum birefringence in the collision of FEL and high-intensity laser pulses is essentially becoming experimentally feasible now. 
At this point, it is worth recalling that in order to simplify the calculation presented in the present notes we limited ourselves to the special scenario, where the FEL and high-intensity laser beams are focussed to the same waist $w$.
Accounting for different waists of the colliding beams, the prospective signal photon number achievable with given laser parameters can be increased even more by optimally choosing the waist parameters of the individual beams \cite{Karbstein:2015xra,Karbstein:2016lby,Karbstein:2018omb,Karbstein:2019bhp}.

\section{Conclusions and Outlook}\label{sec:concls}

In these notes, we have provided a pedagogical introduction to the theoretical study of all-optical signatures of quantum vacuum nonlinearities in strong electromagnetic fields as provided by high-intensity lasers.
More specifically, we have shown in detail how these signatures are analyzed in the vacuum emission picture~\cite{Karbstein:2014fva,Gies:2017ygp}, constituting a very efficient and easy to handle approach for their study in experimentally realistic field configurations \cite{Blinne:2018nbd}. 
As emphasized in the main text, using this approach in combination with a locally constant field approximation of the one-loop Heisenberg-Euler effective Lagrangian, should allow for the accurate quantitative analysis of all-optical signatures of QED vacuum nonlinearity on the one-percent level. 
Though limiting our explicit considerations to a special scenario, namely the head-on collision of two driving laser pulses of the same duration which are focused to the same beam waist, the detailed analysis of this specific example should have familiarized the reader with the basic concepts and techniques needed for the theoretical study of all-optical signatures of quantum vacuum nonlinearity in more general field configurations. 

Correspondingly, we are confident to have provided the reader with a thorough introduction to the topic, allowing her/him to pursue the theoretical study of all-optical signatures of quantum vacuum nonlinearity in his/her favorite electromagnetic field configuration. 

\vspace{6pt} 


\funding{This work has been funded by the Deutsche Forschungsgemeinschaft (DFG) under Grant No. 416607684 within the Research Unit FOR2783/1.}

\acknowledgments{I would like to thank the organizers as well as the participants of the Helmholtz International Summer School (HISS) - Dubna International Advanced School of Theoretical Physics (DIAS-TH) ``Quantum Field Theory at the Limits: from Strong Fields to Heavy Quarks'' for making the 2019 edition of the summer school in Dubna a very inspiring and enjoyable event.
Moreover, I am grateful to Elena Mosman for a careful and critical reading of the manuscript, as well as to Stephan Fritzsche for interesting discussions and a fresh beer~\cite{Fritzsche:2019} to atomic structures, processes and cascades.}

\conflictsofinterest{The authors declare no conflict of interest.} 

\reftitle{References}



\end{document}